# Essential role of liquid phase on melt-processed GdBCO single-grain superconductors

Xiongfang Liu [a], Xuechun Wang [a], Jinyu He [a], Yixue Fu [a], Xinmao Yin [a], Chuanbing Cai [a], Yibing Zhang [*a], Difan Zhou [*a]

[a] Shanghai Key Laboratory of High Temperature Superconductors, Department of Physics, Shanghai University, Shangda Road 99, 200444 Shanghai, China
[*] Corresponding author
E-mail address: zyb_shu@shu.edu.cn (Yibing Zhang).
              dz286@shu.edu.cn (Difan Zhou).

**Abstract:**

RE–Ba–Cu–O (RE denotes rare earth elements) single-grain superconductors have garnered considerable attention owning to their ability to trap strong magnetic field and self-stability for maglev. Here, we employed a modified melt-growth method by adding liquid source (LS) to provide a liquid rich environment during crystal growth. It further enables a significantly low maximum processing temperature ($T_{max}$) even approaching peritectic decomposition temperature. This method was referred as the liquid source rich low $T_{max}$ (LS+$LT_{max}$) growth method which combines the advantage of Top Seeded Infiltration Growth (TSIG) into Top Seeded Melt-texture Growth (TSMG). The LS+$LT_{max}$ method synergistically regulates the perfect appearance and high superconducting performance in REBCO single grains. The complementary role of liquid source and low $T_{max}$ on the crystallization has been carefully investigated. Microstructure analysis demonstrates that the LS+$LT_{max}$ processed GdBCO single grains show clear advantages of uniform distribution of $RE^{3+}$ ions as well as RE211 particles. The inhibition of Gd211 coarsening leads to improved pining properties. GdBCO single-grain superconductors with diameter of 18 mm and 25 mm show maximum trapped magnetic field of 0.746 T and 1.140 T at 77 K. These trapped fields are significantly higher than those of conventional TSMG samples. Particularly, at grain boundaries with reduced RE211 density superior flux pinning performance has been observed. It indicates the existence of multiple pinning mechanisms at these areas. The presented strategy provides essential LS+$LT_{max}$ technology for processing high performance single-grain superconductors with improved reliability which is considered important for engineering applications.

## 1. Introduction

In 1986, copper oxides are known to exhibit superconductivity above exceeding liquid nitrogen temperature by Bednorz and Müller[1]. Family of high-temperature superconductors (HTSs) welcomed a new member Y-Ba-Cu-O[2]. Subsequently, a series of high-temperature superconductors with transition temperature around 92 K were obtained by replacing Y with rare earth elements Gd, Sm, Nd, Eu, Yb, *etc*. According to different application scenarios, RE–Ba–Cu–O HTSs can be processed into single crystals, large quasi-single grains, thin films, coated conductors (CCs) and devices based on heterojunctions. The core commonality is the crystallization/texturization of REBCO since the superconductivity mainly exists on the two-dimensional Cu-O planes. Among these configurations, single-grain HTSs show strong ability for trapping magnetic flux [3-8] and self-stability for maglev which make them be widely used for trapped-field magnets, superconducting magnetic resonance microscopes, superconducting magnetic lens, superconducting levitation systems and superconducting flywheel energy storage[9-13].

There are presently two main growth methods for fabricating single-grain $REBa_2Cu_3O_{7-\delta}$ materials. One is Top Seeded Melt-texture Growth (TSMG) method [14-18], the TSMG aimed at melting and recrystallizing $REBa_2Cu_3O_{7-\delta}$ (RE123) phase. The other is Top Seeded Infiltration Growth (TSIG) method [19-22] where Ba-Cu-O liquid source permeates into $RE_2BaCuO_5$ (RE211) pre-forms due to the siphon effect. Therefore, TSIG compared with TSMG allows a lower maximum temperature which benefits the stability of the seeding crystals and the inhibition of RE211 coarsening[19]. Considerably modification has been made for the liquid source in TSIG.

Referring to the TSIG, a modification to TSMG by providing additional liquid source during growth has been widely adopted. Cardwell *et al* [23] concluded that appropriate liquid-rich phase may yield reliable REBCO single grains growth without formation of sub-grains at the edge and bottom. Zhou *et al* [24, 25] demonstrated that Y123 pressed pellet can function as liquid source supplement to optimize the growth of GdBCO bulk superconductors. Liquid phase also plays essential role in REBCO thin film growth. Yao *et al*[26, 27] reported successful growth of well-crystalline REBCO films by liquid phase epitaxy. Particularly, recent developed transient liquid assisted growth (TLAG) [28] of REBCO CCs has drawn great attention. The transient liquid forming during growth changes the solid reaction into liquid reaction greatly facilitating the dispersion of RE ions. As a result, ultra-fast growth rates and uniform growth have been realized. TLAG technology has been attempted by the Chemical Solution Deposition (CSD)[29] and Pulsed Laser Deposition (PLD)[30] processing for preparing long REBCO tapes, and could be a game changing technology. It urges us to re-visit the effects of the liquid rich environment during TSMG processing.

In this work, we added Y123 layer as liquid source (LS) supplement and adopted low maximum processing temperature ($LT_{max}$) to fabricate GdBCO single-grain superconductors. This modified melt-growth method was referred as the LS+$LT_{max}$ growth method (with Y123 as LS and $T_{max,\ LS+LT_{max}}$ of 1048 °C). The Ba-Cu-O liquid phase comes from the decomposition of Y123 pellets and then infiltrates into REBCO matrix leading to liquid enriched surroundings. The LS enables successful growth of REBCO single grains with reduced maximum temperature close to the peritectic decomposition temperature ($T_p$). The $LT_{max}$ growth further inhibits coarsening of RE211 inclusions[4, 31]. Owning to the dual contribution of LS and $LT_{max}$, the single-grain superconductors presented superior superconducting performance. The morphological characteristics and superconducting properties have been systematically investigated. Fig.1 shows the principle and procedures of the LS+$LT_{max}$ process and the comparison to the conventional TSMG technology.

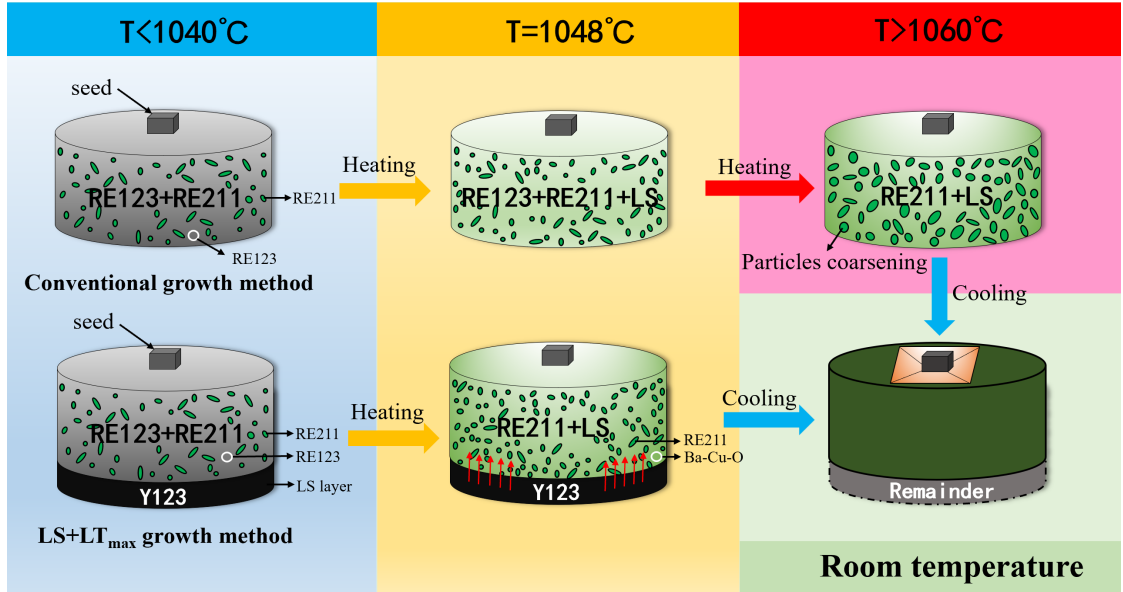

Fig. 1. Influential phenomena of heat-treating REBCO superconducting crystal states at different temperatures by using the Conventional growth method and the LS+LT$_{max}$ growth method.

## 2. Experimental details

### 2.1 Fabrication of LS+LT$_{max}$ GdBCO single grains

Melt-processed GdBCO single-grain superconductors were prepared using high-purity home-made precursor powders of GdBa$_2$Cu$_3$O$_{7-\delta}$(Gd123), Gd$_2$BaCuO$_5$(Gd211) and YBa$_2$Cu$_3$O$_{7-\delta}$(Y123), which were subjected to solid-state reaction twice after weighing Gd$_2$O$_3$, BaCO$_3$ and CuO compounds in nominal composition. The precursor powders of Gd123, Gd211 and CeO$_2$ additive were mixed using a ball-milling technique with a composition of (Gd123+35%mol Gd211) + 0.5 wt% CeO$_2$. The addition of CeO$_2$ was used to control the coarsening of Gd211 particles[32]. The mixed compound and Y123 were uniaxially pressed into cylinders of 20 mm and 28 mm diameter, respectively, with different thicknesses. Y123 was used as a liquid source slice placed beneath the pre-forms of the mixed compound. A generic seed (Sm123/Nd123) was placed on the top surface of the pre-forms at room temperature (Cold-seed technology) to promote heterogeneous nucleation in the molten liquid[33]. Then, the entire arrangement was placed onto the Yb$_2$O$_3$ substrate into the box furnace, detailed arrangement and thermal profile as shown in Fig. 2(a). The maximum value of processing temperature was set to 1048 °C ($T_{max, LS+LT_{max}}$ = 1048 °C), which is very close to the peritectic decomposition temperature. Next, it with a slow cooling rate to obtain single grain. The annealing process was carried out in flowing pure oxygen, the sample was heated to 400 °C−420 °C and held for 240 h, eventually it was cooled down to room temperature.

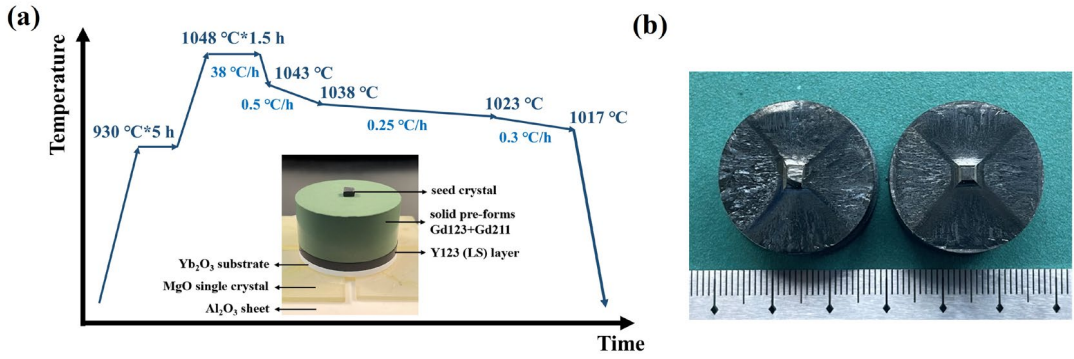

Fig. 2. (a) Temperature profile of the LS+LT$_{max}$ growth method in processing of GdBCO single grain. The inset shows the entire arrangement of the pre-form geometry structure. (b) Top view of the 25mm diameter LS+LT$_{max}$ GdBCO single grains seeded by a small Sm123/Nd123 crystal square.

### 2.2 Exploration of LS+LT$_{max}$ growth mechanism

To determine the specific conditions of the LS+LT$_{max}$ growth method, one needs to advance in the understanding of the mechanism of the LS+LT$_{max}$ growth method, which we tackled in this study with two experiments, namely "Powders quench experiment" and " Bulk critical temperature experiment ", respectively.

Firstly, to analyze the role of LS in the peritectic reaction of Gd123, a "Powders quench experiment" was conducted. The temperature profile of this experiment was set as follows: direct ramped up to peritectic decomposition temperature, held for 1 hour, and rapid cooled to room temperature. The reaction products of Gd123 powders and Gd123+Ba-Cu-O mixture powders were analyzed using X-ray Diffraction (XRD). Secondly, the "Bulk critical temperature experiment" was conducted to further understand the importance of the LS in growth of REBCO single grain. Temperature profile for the "Bulk critical temperature experiment" is shown in the Fig. 6(c): the maximum processing temperature was set equal to the peritectic decomposition temperature ($T_{max} = T_p$) and held for 1.5 hours. The initial supercooling temperature was less than 5 °C, and the cooling rate was 0.2-0.3 °C/h from $T_{grow1}$ to $T_{grow2}$.

### 2.3 Superconducting properties of LS+LT$_{max}$ GdBCO single grains

The superconducting properties of single-grain superconductors are a key factor in determining its potential for practical applications. The maximum trapped magnetic field ($B_{tr,max}$), trapped magnetic flux density distribution, superconductive transition temperature ($T_c$), local critical current density ($J_c$) are essential parameters for assessing the superconducting properties of a single-grain superconductors.

The LS+LT$_{max}$ GdBCO single grains were polished smooth, then their $B_{tr,max}$ and trapped magnetic flux density distribution at 77 K were measured. This process involved cooling the single grains to 77 K in an applied magnetic field of 1.5 T and maintaining the applied magnetic field for 15 minutes to stabilize the magnetic flux[4]. The $B_{tr,max}$ was then confirmed using a hand-held Hall probe for an accurate value, and the complete trapped magnetic flux density distribution profile was subsequently measured using a scanning Hall probe. Quantum Design-Physics Property Measurement System (PPMS) was used to measure Magnetic-Hysteresis (*M-H*) loops at 77 K and measure $T_c$ curves from 80 K to 100 K. Several slices (size about 2 mm×2 mm×0.5 mm, labelled as C1, C2, GB1, GB2, B1, B2 plotted in Fig. 3(a)) were selected from LS+LT$_{max}$ GdBCO single grain to measure *M-H* loops. Then $J_c$ was calculated using the extended Bean model[34]. Another set of slices (labelled as C1, C2, C3, B1, B2 plotted in Fig. 3(b)) were selected for $T_c$ testing in the LS+LT$_{max}$ GdBCO single grain.

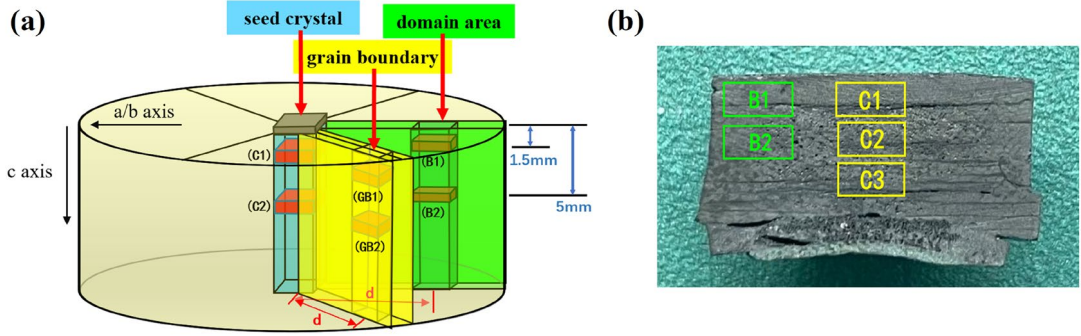

Fig. 3. (a) The positions of six slices (C1, C2, GB1, GB2, B1, B2) in LS+LT$_{max}$ GdBCO single grain for *M-H* and FE-SEM testing. These slices were located at center, grain boundaries and domain areas, respectively. The position of C1, GB1, B1 is 1.5 mm from top surface and the position of C2, GB2, B2 is 5 mm from top surface. GB1, GB2 and B1, B2 are at the same distance from the center. (b) Depicts the cross-section of the LS+LT$_{max}$ GdBCO single grain, with slices (C1, C2, C3, B1, B2) taken from regularly selected locations.

**2.4 Micro-analysis of LS+LT$_{max}$ GdBCO single grains**

Micro-analysis act as essential role in determining superconductivity and mechanical properties of REBCO single-grain superconductors. Size and distribution of inclusions, holes, cracks, growth direction, grain boundaries, ion concentration can be observed by Micro-analysis[35]. Micro-analysis provided valuable insights into the microstructure and composition of LS+LT$_{max}$ GdBCO single grain, helping to uncover the reasons for their high superconducting performance.

A Field Emission Scanning Electron Microscope (FE-SEM) was used to analyze the GdBCO single grain for microstructure. The average size and local distribution of RE211 particles in the LS+LT$_{max}$ GdBCO single grain and the Conventional GdBCO single grain were quantitatively observation by using ImageJ software[4]. It is noteworthy that the set of slices used for FE-SEM testing are the same as those used for $J_c$ testing, as shown in Fig. 3(a). In addition, given that the status of Gadolinium ions is directly related to the growth of the GdBCO single grain, therefore, a "Bulk quench experiment" (Gd123 + Gd211 as precursor powders, heated to $T_{max}$ = 1048 °C and maintained for 1.5 hours, then rapidly cooled to room temperature, with no seed crystal placed) was conducted and Energy Dispersive Spectroscopy (EDS) was employed to investigate the concentration and distribution of Gadolinium ions at the growth front in LS and non-LS samples.

# 3、Result and discussion

## 3.1 Morphological characteristics and trapped field ($B_{tr}$)

Morphological characteristics and trapped field ($B_{tr}$) of REBCO single grains are significant and intriguing observations. In light of the crystal growth model, the interfacial movement of RE123 occurs simultaneously with the coarsening of RE211 particles in molten liquid[25, 36-38]. Based on our observations, it is difficult to reach strong magnetic flux pinning ability by using the Conventional growth methods (without LS, high $T_{max}$), owing to the large size of the RE211 particles. However, by utilizing the LS+LT$_{max}$ growth method, low maximum processing temperature (LT$_{max}$) can set a barrier for coarsening of RE211 particles, while LS can promote atomic diffusion and also contribute to prevent coarsening of RE211 particles[23, 39].

Our experimental results show that all $B_{tr}$ images exhibit a single-peak model. However, the magnetic flux distribution schematic diagrams and $B_{tr,max}$ between the Conventional single grains and the LS+LT$_{max}$ single grains present a marked difference. In the Conventional single grain with diameter of 18mm [Figs. 4(a)(b)], the distribution schematic diagram of magnetic flux shows a polygonal shape, has no symmetry. Besides, its $B_{tr,max}$ is only 0.445 T. However, the two-dimensional magnetic flux distribution schematic diagram of the LS+LT$_{max}$ single grain displays multiple perfect concentric positive circles and its three-dimensional magnetic flux distribution schematic diagram is plump and smooth. Furthermore, the $B_{tr,max}$ of the LS+LT$_{max}$ single grain with diameter of 18mm is 0.746 T [Figs. 4(c)(d)]. Those indicate that the LS+LT$_{max}$ single grain provides a more perfect trapped magnetic flux. Meanwhile, distinct cases are observed between the LS+LT$_{max}$ single grain and the Conventional single grain, both of which have a 25mm diameter. The $B_{tr,max}$ of the Conventional single grain is 0.710 T with irregular magnetic flux distribution [Figs. 4(e)(f)], while the $B_{tr,max}$ of the LS+LT$_{max}$ single grain is upgraded by 60.56 %, reaching a value of 1.140 T [Figs. 4(g)(h))]. Therefore, we conclude that the LS+LT$_{max}$ growth method can form well-textured REBCO single grains, optimize the trapped magnetic flux distribution inside the bulks and improve the overall trapped magnetic field performance by controlling the coarsening of RE211 particles.

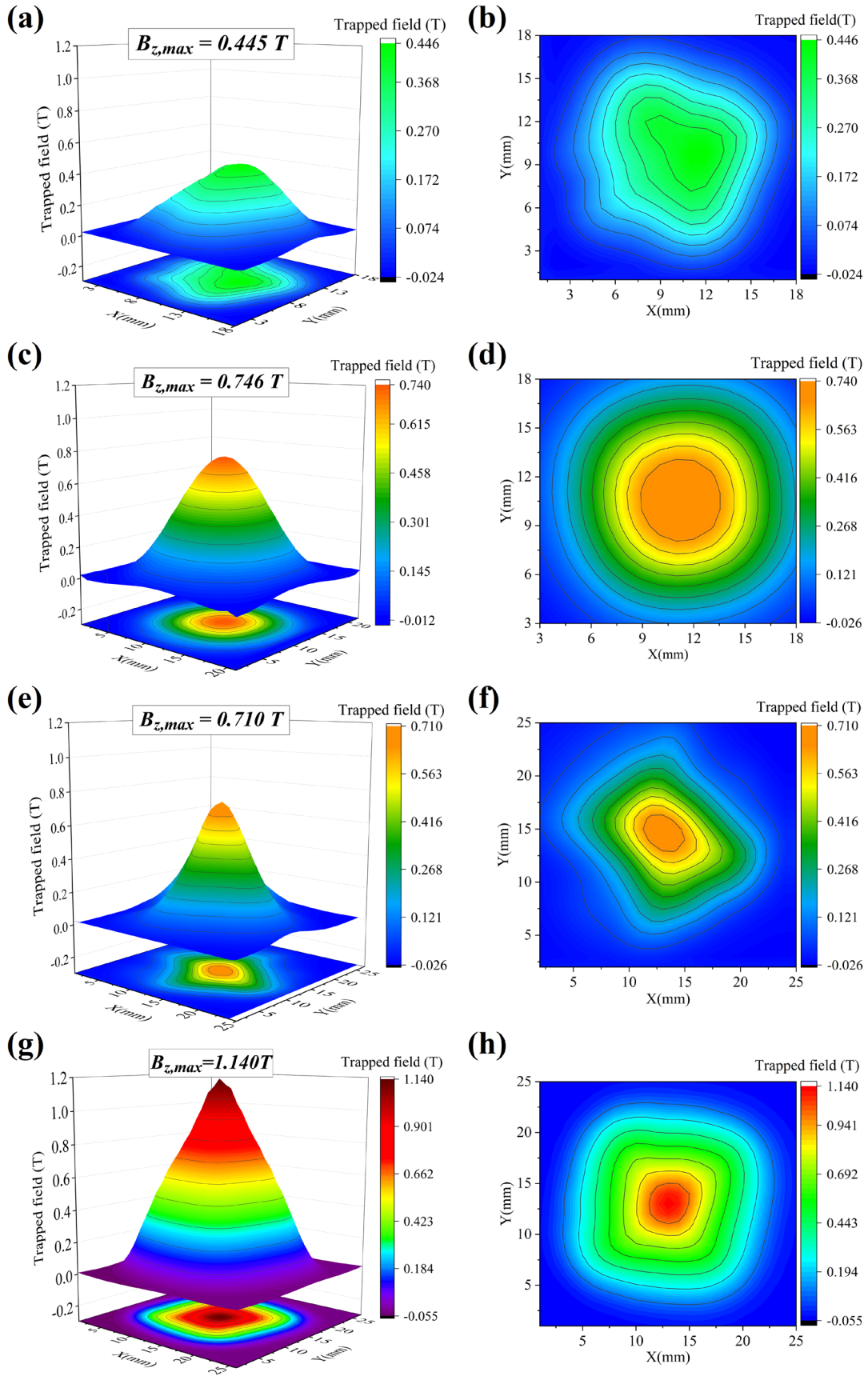

Fig. 4. Three-dimension (left) and two-dimension (right) magnetic flux distribution schematic

diagrams of GdBCO single grains. (a)(b) Conventional single grain with diameter of 18mm; (c)(d) LS+LT$_{max}$ single grain with diameter of 18mm; (e)(f) Conventional single grain with diameter of 25mm; (g)(h) LS+LT$_{max}$ single grain with diameter of 25mm.

In order to strengthen our viewpoints about the LS+LT$_{max}$ growth method can fabricate well-textured single-grain superconductors, we produced multiple LS GdBCO-Ag and non-LS GdBCO-Ag samples, and established three sets of comparison samples for this experiment. The GdBCO-Ag system underwent a range of maximum processing temperatures, including $T_{max,Gd-Ag}$, $T_{max,Gd-Ag}$+5 °C and $T_{max,Gd-Ag}$+10 °C.

Figs. 5(a)(c)(e) show the spontaneous nucleation behaviors of three non-LS GdBCO-Ag samples, which we found increased as the maximum processing temperature decreased. Interestingly, the spontaneous nucleation behaviors are refined and clean surfaces are realized after adding the LS [Figs. 5(b)(d)(f)]. Subsequently, we conducted measurements of the $B_{tr}$ performance of LS GdBCO-Ag samples. The $B_{tr,max}$ increases monotonically (0.552 T − 0.630 T − 0.706 T) as the temperature decreasing, providing a new evidence on how LT$_{max}$ and a rich LS synergistically inhibit spontaneous nucleation behavior and enhance trapped magnetic field performance. The optimized LT$_{max}$ and LS are able to effectively control the size and content of the Gd211 particles in final microstructure, which intimately involves in effective flux pinning and transport critical current ($J_c$).

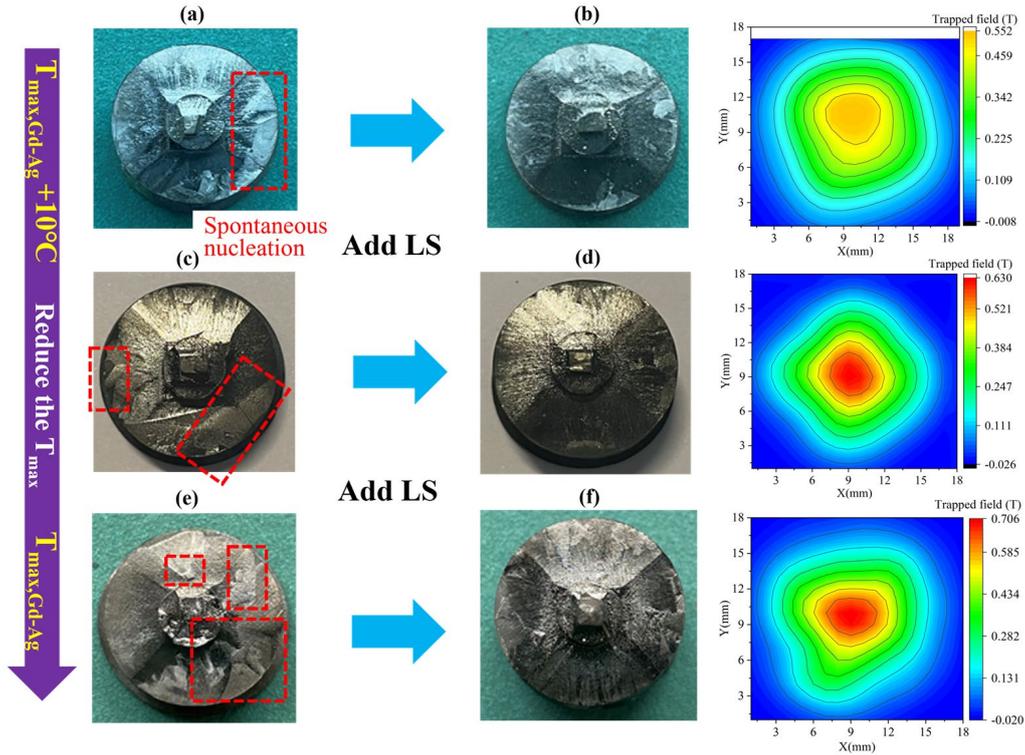

Fig. 5. Schematic illustrations of the morphological characteristics and $B_{tr}$ in GdBCO-Ag system. (a)(c)(e) The spontaneous nucleation sites gradually increase with decreasing the maximum processing temperature in non-LS samples. (b)(d)(f) The addition of LS refines the spontaneous nucleation behaviors and obtains relatively clean surface. More excellent $B_{tr}$ performance can be observed as the maximum processing temperature decreasing in LS GdBCO-Ag samples.

## 3.2 The mechanism of LS+LT$_{max}$ growth method

As mentioned above, the LS+LT$_{max}$ growth method to be particularly attractive due to its advantages. The LS+LT$_{max}$ growth method, which involves wet liquid surroundings[40] that dissolve more solid particles by permeation of Ba-Cu-O liquid source (comes from decomposition of Y123) and relies on a strong chemical diffusion driving force[41] to increase the transport rate in the molten liquid. Moreover, due to its high-diffusion rate, The LS sintering method also facilitates faster sintering reactions and lower the sintering temperature[25, 36-38], enabling a significantly low maximum processing temperature even approaching peritectic reaction temperature. To further explore the mechanism of LS+LT$_{max}$ growth method, two experiments: the "Powders quench experiment" and the "Bulk critical temperature experiment" have been carried out. Based on analysis above, Ba-Cu-O powders were directly added in these two experiments for forming enriched liquid source surroundings and participating in peritectic reaction. That will change the peritectic decomposition reaction temperature ($T_p$) of RE123 phase and form a perfect peritectic reaction[41].

In the results of our experiment, the XRD pictures [Fig. 6(a)] of the "Powders quench experiment" show that the reaction product of Gd123 powders is almost identical to itself and indicate that no decomposition reaction has occurred in Gd123 powdesr when the $T_{max}$ equal to its $T_p$. However, the XRD pictures indicate that the reaction product of the Gd123+Ba-Cu-O mixture powders have a significant concentration of the Gd211 phase. This illustrates that Gd123+Ba-Cu-O mixture powders indeed undergo a peritectic decomposition reaction when the $T_{max}$ equal to its $T_p$. Combined with the data of DTA testing (Fig. 6(b)), the actual $T_p$ of the Gd123+BaCuOx mixture powders is 1042 °C ($T_{p1}$ = 1042 °C) and the actual $T_p$ of the Gd123 powders is 1044 °C ($T_{p2}$ = 1044 °C). $T_{p1}$ is 2 °C lower than $T_{p2}$. This indicates that the LS can effectively reduce the actual peritectic reaction temperature of Gd123 and decrease the solid-liquid line in the phase diagram of Gd123[42]. It should be mentioned that in these experiments, only a small amount of LS powders (5wt% Ba-Cu-O) were added, less than the amount of LS added in the actual bulk samples.

Experimental outcomes of "Bulk critical temperature experiment" [Fig. 6(c)] demonstrate that the bulk without LS hardly realize growth when the $T_{max}$ equal to its $T_p$. It has been observed that the top surface of non-LS bulk is covered with heterogeneous nucleation sites. However, with the help of LS, another bulk is able to successfully grow one domain when the $T_{max}$ equal to its $T_p$. This suggests that the function of LS during the growth process is to accelerate the diffusion of atoms, lower the $T_p$, and promote the melting of the Gd123 phase[41]. Therefore, it is reasonable to believe that the enriched LS can enable a significantly low maximum processing temperature even approaching peritectic reaction temperature, then growing such wonderful single grains by employing LS+LT$_{max}$ growth method.

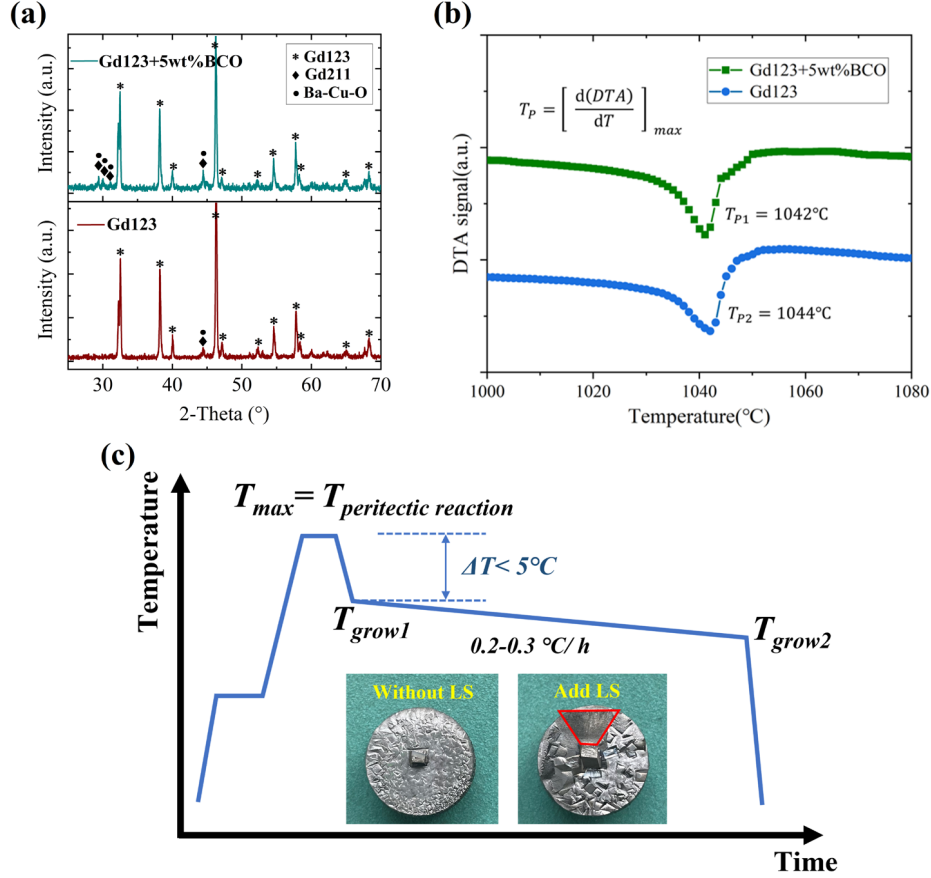

Fig. 6. (a) Analysis of the reaction product of Gd123 powders and Gd123+Ba-Cu-O mixture powders on "Powders quench experiment" by using XRD. (b) Analysis of actual $T_p$ of Gd123 powders and Gd123+Ba-Cu-O mixture powders by using DTA respectively. (c) Temperature profile of "Bulk critical temperature experiment", the insets shows the top surface of two bulks with or without LS.

### 3.3 Superconducting properties of LS+LT$_{max}$ GdBCO single grains: $J_c$ and $T_c$

$J_c$ and $T_c$ are the fingerprint of superconducting properties. A few slices [Fig.3(a)] were used to measure $J_c$ at 77 K. According to the push/trapped theory of RE211 particles, the position where near the seed crystal contains the small-sized RE211 particles, thus producing more significant interfacial defects, leading to a stronger flux pinning property and a higher $J_c$ value at lower fields [43-46]. However, the penetration of LS causes more connectivity of the liquid and increases the liquid diffusion flux[41]. Thereby, on the one hand, the LS amplifies the pushing effect of RE211[47-49] and decreases the $J_c$ value near the seed crystal. On the other hand, LT$_{max}$ make contribution to excellent $J_c$ behaviors by preventing coarsening of RE211particles in the whole LS+LT$_{max}$ single grain.

Fig. 7(a) indicates that $J_c$ of C1/C2 regions indeed exhibits small $J_c$ value in the LS+LT$_{max}$ single grains at low field. The main reason is less Gd211 particles in C1/C2 regions due to the amplified pushing effect of Gd211 by LS. Therefore, less pining center in the C1/C2 region. Meanwhile, the $J_c$ increases as the distance from the center position increasing at low fields. This is attributed to the presence of more Gd211 particles outside of the center. In addition, the maximum value of $J_c(0)$ are $5.1 \times 10^4 A/cm^2$ at grain boundary (GB1) and $5.0 \times 10^4 A/cm^2$ at domain area (B1). Apart from that, The "Peak value effect" can be clearly identified at grain boundaries, this is a manifestation of the Gd–Ba substitution which can become the flux pining center at high field[50, 51].

The set of slices reveals to role of grain boundary flux pining mechanism so as to complement existing studies of flux pining effect of HTSs [52]. The $J_c(0)$ of grain boundaries are higher than domain areas (GB1>B1; GB2>B2) in the same ab plane, as seen in the inset of Fig. 7(a). The enhanced effective pinning function at grain boundaries originates from special defects, such as dislocations, twins and stack faults[53-57], which suppress vortex channeling at 77K. Besides, Zerweck [58] have supported electron scattering mechanism at grain boundaries. He pointed out that grain boundaries and magnetic flux lines have an attractive interaction, thus strengthening the flux pining property at grain boundaries.

The result of superconducting magnetic susceptibility measurement (*M-T*) presents in Fig. 7(b) and demonstrates that the LS+LT$_{max}$ GdBCO single grain exhibits excellent superconducting performance. All of the slices [Fig. 3(b)] have superconducting transition temperature ($T_c$) above 94.1 K. Additionally, the *M-T* curves of different positions are similar, with a sharp transition and a transition width of less than 3 K. This is due to the homogeneity of the materials, which is induced by the enrichment of LS.

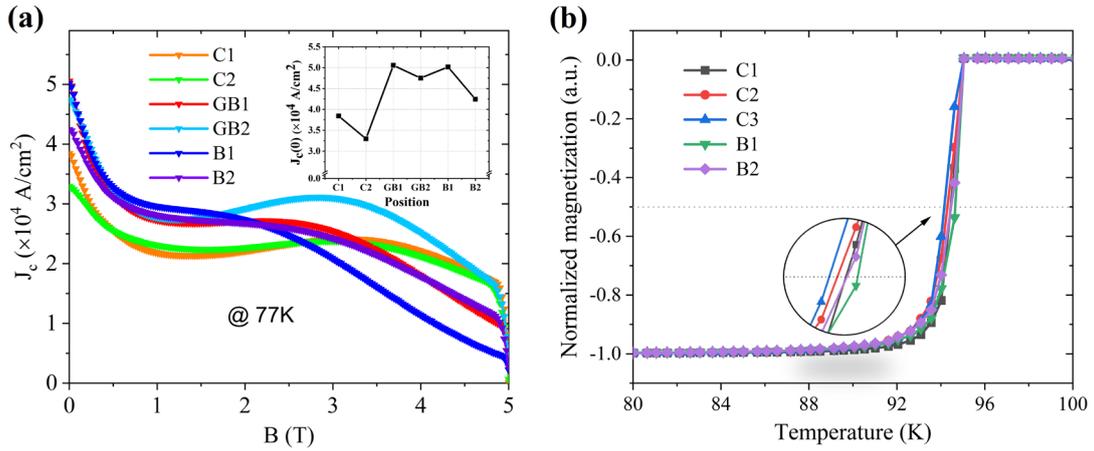

Fig. 7. (a) $J_c$–$B$ curves of LS+LT$_{max}$ GdBCO single grain with different positions of slices at 77 K. (b) The superconducting transition temperature ($T_c$) was evaluated at various positions within the LS+LT$_{max}$ GdBCO single grain.

### 3.4 Microstructure analysis of LS+LT$_{max}$ GdBCO single grains: FE-SEM and EDS

More detailed research demand to utilize the FE-SEM and EDS for microstructure analysis. During the preparation of the slices for microstructure analysis, it is necessary to polish them with metallographic sandpaper. As a result, all slices exhibit varying degrees of mechanical scratches. However, it is pleased that these mechanical scratches do not impede the observation of the distribution and size of the Gd211 particles embedded within the slices. Quantitative analysis of size and distribution of the Gd211 particles at regions C1, C2, GB1, GB2, B1, B2 by using the ImageJ software.

The microstructure of the Conventional single grains is depicted in Fig. 8(a). The Gd211 particles manifest an increasing trend in size and population from center to peripheral regions. This trend is in agreement with the push/trapped theory of RE211 particles[47-49]. The coarsening and coalescence of Gd211 particles are prominently discernible within matrix of the Conventional single grain, which will ultimately lead to a marked decrease in $B_{tr,max}$ and $J_c$ behaviors. Fig. 9(a) depicts the microstructure of the LS+LT$_{max}$ single grain and reveals that the Gd211 particles are distributed relatively independently within the matrix and seldom observed in clusters. Owing to the enhanced pushing effect exerted by the LS, the C1 and C2 regions embed a lower population of Gd211 particles in the LS+LT$_{max}$ single grain relative to the Conventional single grain. The Gaussian distribution curves of size of Gd211 particles in the Conventional single grain and LS+LT$_{max}$ single grain are portrayed by the red lines in Figs. 8(b) and 9(b).

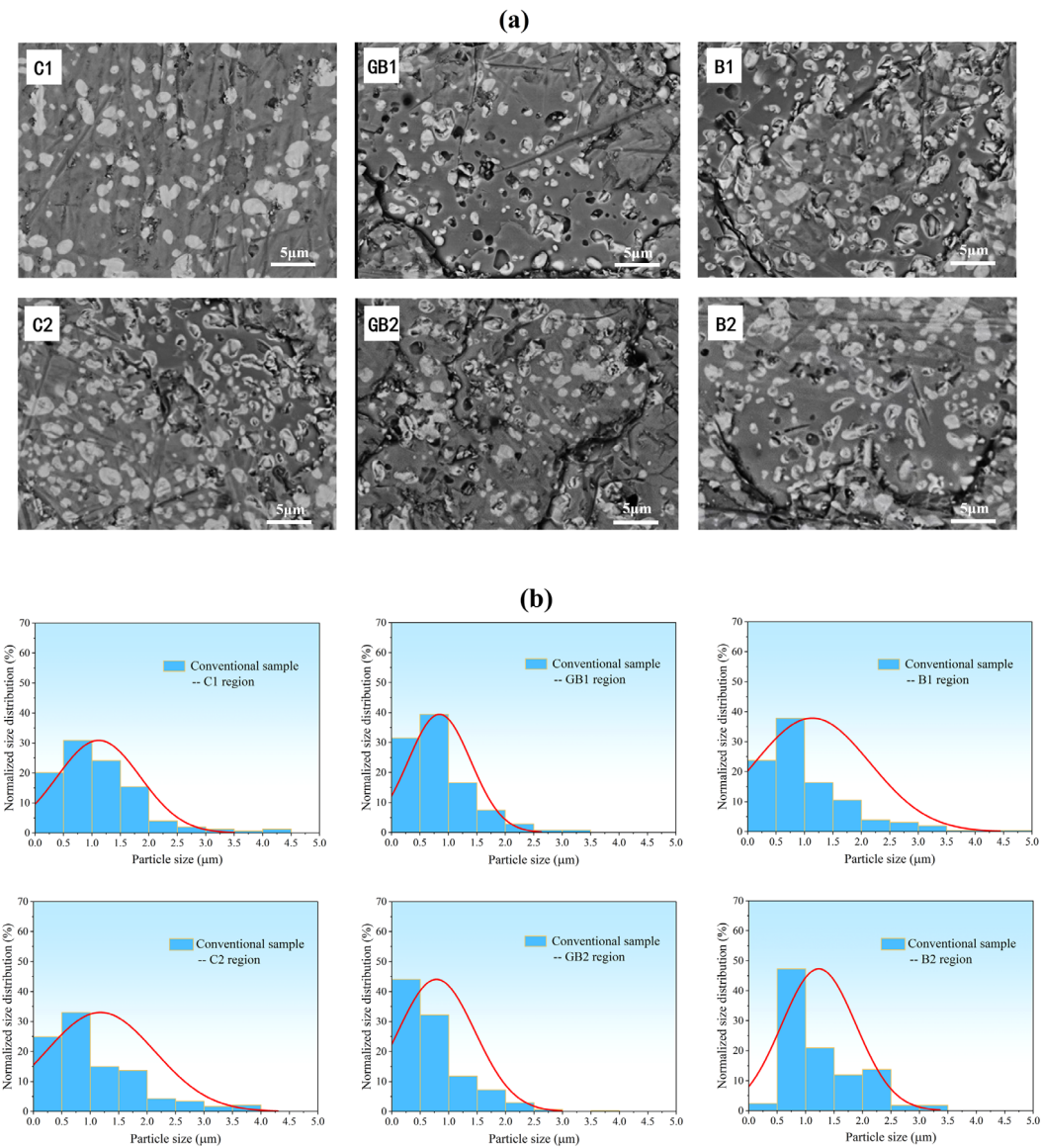

**Fig. 8.** (a) Representative FE-SEM micrographs and (b) Histogram of the diameter and number of Gd211 particles at selected regions in the Conventional single grain.

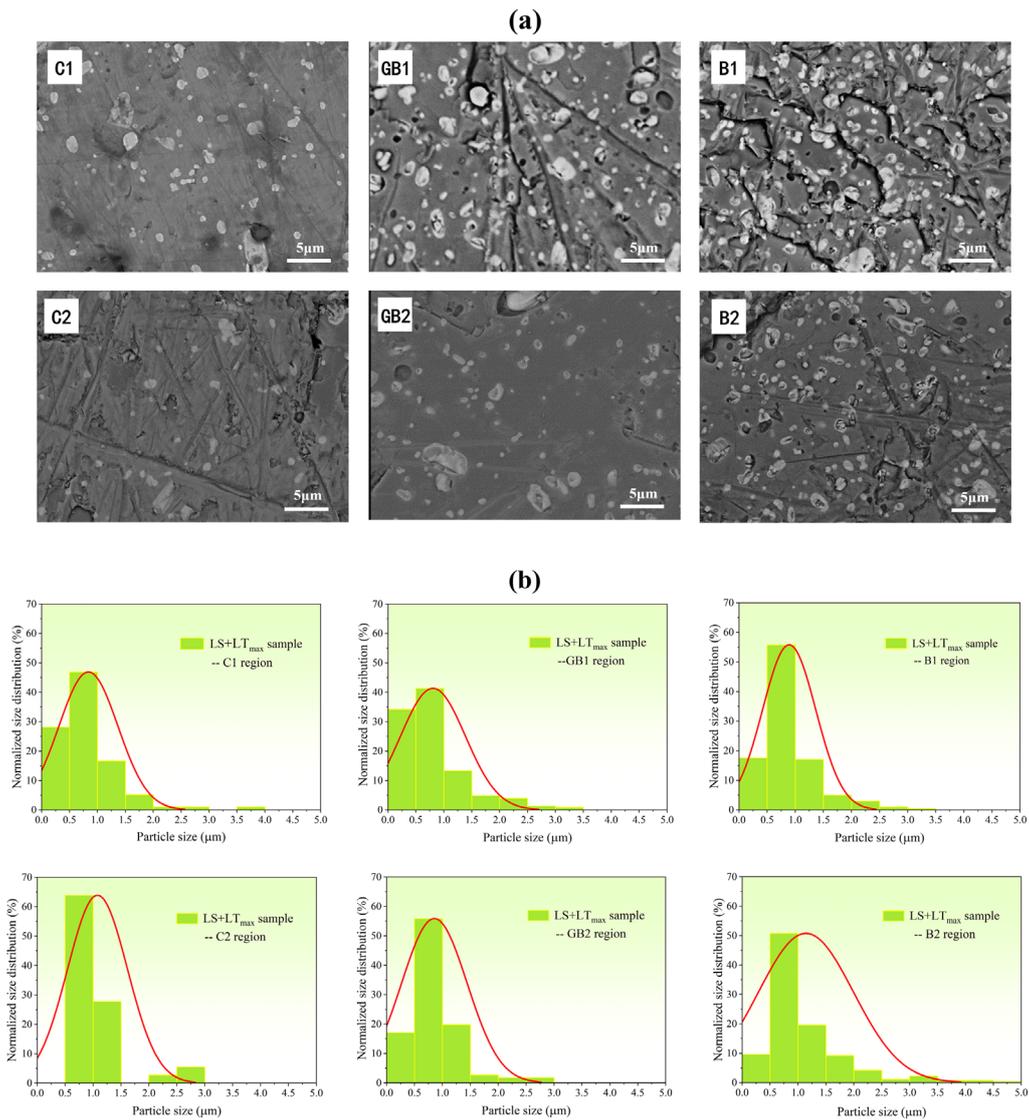

**Fig. 9.** (a) Representative FE-SEM micrographs and (b) Histogram of the diameter and number of Gd211 particles at selected region in LS+LT$_{max}$ single grain.

Further statistical analysis studies (Fig. 10) are conducted on the Conventional single grain and the LS+LT$_{max}$ single grain about their average size and number of Gd211 particles at different regions. The average size of Gd211 particles at different regions of the Conventional single grain exhibits considerable variation, as demonstrated by the line graphs that portray the average size of Gd211 particles. In contrast, the average size of Gd211 particles at various regions of the LS+LT$_{max}$ sample displays minor variation and smaller size of Gd211 particles. This can be considered that the rich LS enters into interior of the single grains, reducing the temperature gradient at various regions and promoting the thermodynamic equilibrium. Meanwhile, the LT$_{max}$ suppresses the coarsening of Gd211 particles. Besides, the LS+LT$_{max}$ single grain exhibits a fewer Gd211 particles compared to the Conventional single grain, as demonstrated by the histograms that portray the number of Gd211 particles. It can be attributed that the LS enhances the push effect of Gd211 particles and the LT$_{max}$ plays a dominant role in alleviating the macro-segregation of Gd211 particles, which will make more Gd211 particles become effective flux pinning centers in the LS+LT$_{max}$ single grain.

Of paramount significance, based on comprehensive statistical analysis, a significant discovery is made regarding the difference of number of Gd211 particles in the domain areas and grain boundaries, which bears significant implications for the field of materials science. In both the LS+LT$_{max}$ single grain and the Conventional single grain, a decreasing trend in number of Gd211 particles at the grain boundaries compared to the domain areas can be observed. Nevertheless, $J_c$ has demonstrated that the grain boundaries process excellent magnetic flux pinning properties compare to domain areas [Fig. 7(a)]. This finding provides new evidence that the RE211 particles are an important factor, but it is not the sole determinant of $J_c$. There exist multiple pinning mechanisms at grain boundaries, such as dislocations, twins and stack faults[53-57]. Those defects can effectively control the motion of magnetic flux vortex and reduce energy dissipation, thus enhancing $J_c$ value[59, 60].

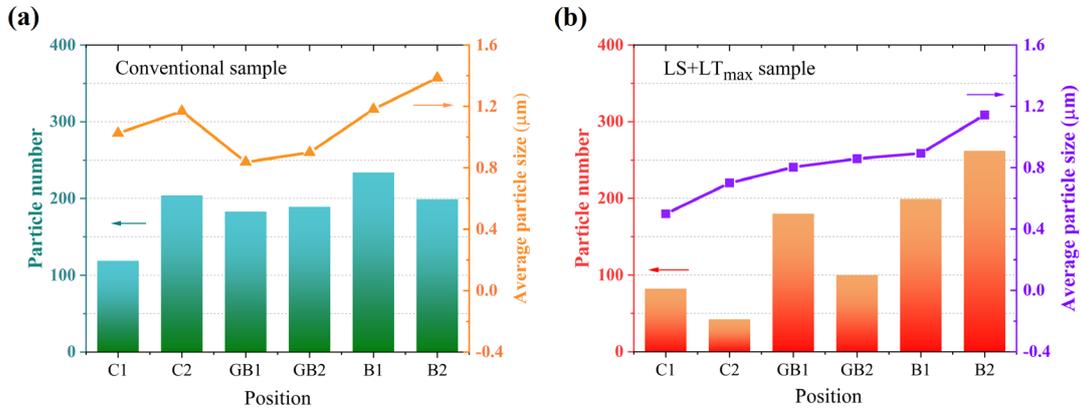

Fig. 10. Histograms of the average particle size of Gd211 inclusions and line graphs of particle number of Gd211 inclusions in (a) the Conventional single grain and (b) the LS+LT$_{max}$ single grain.

The established growth model of REBCO single grains has garnered widespread support from researchers in this field[61, 62]. At the growth front, the RE ions that are decomposed from RE211 particles combine with the Ba-Cu-O liquid source, leading to the formation of a textured RE123 phase under the guidance of a seed crystal. In our opinion, the quality of the single grain is contingent upon the concentration and distribution of RE ions which present at the growth front. A higher concentration and uniform distribution of RE ions are indicative of a well-textured single grain. Our experimental finding reveals that the cross-section photograph of the non-LS sample displays a considerable amount of undecomposed Gd211 precursor powders [Fig. 11(d)]. Additionally, the EDS mappings of various positions suggest a lower concentration and greater inhomogeneity of Gd ions [Figs. 11(a-c)], which could lead to random nucleation. However, the cross-section photograph of the LS sample displays a lower amount of remaining Gd211 precursor powders [Fig. 11(h)], which indicates a more thorough decomposition of the Gd211 precursor powders. Furthermore, the EDS mappings of different positions demonstrate a higher concentration and more uniform distribution of Gd ions [Figs. 11(e-g)]. We believe that the enriched LS will facilitate the global and homogeneous transport of RE ions during the melting process. This will lead to a more uniform growth and a reduction in random nucleation, thus resulting in a more uniform distribution of the RE211 particles in final microstructure. These factors will contribute to the excellent superconducting performance of the REBCO single-grain superconductors as well.

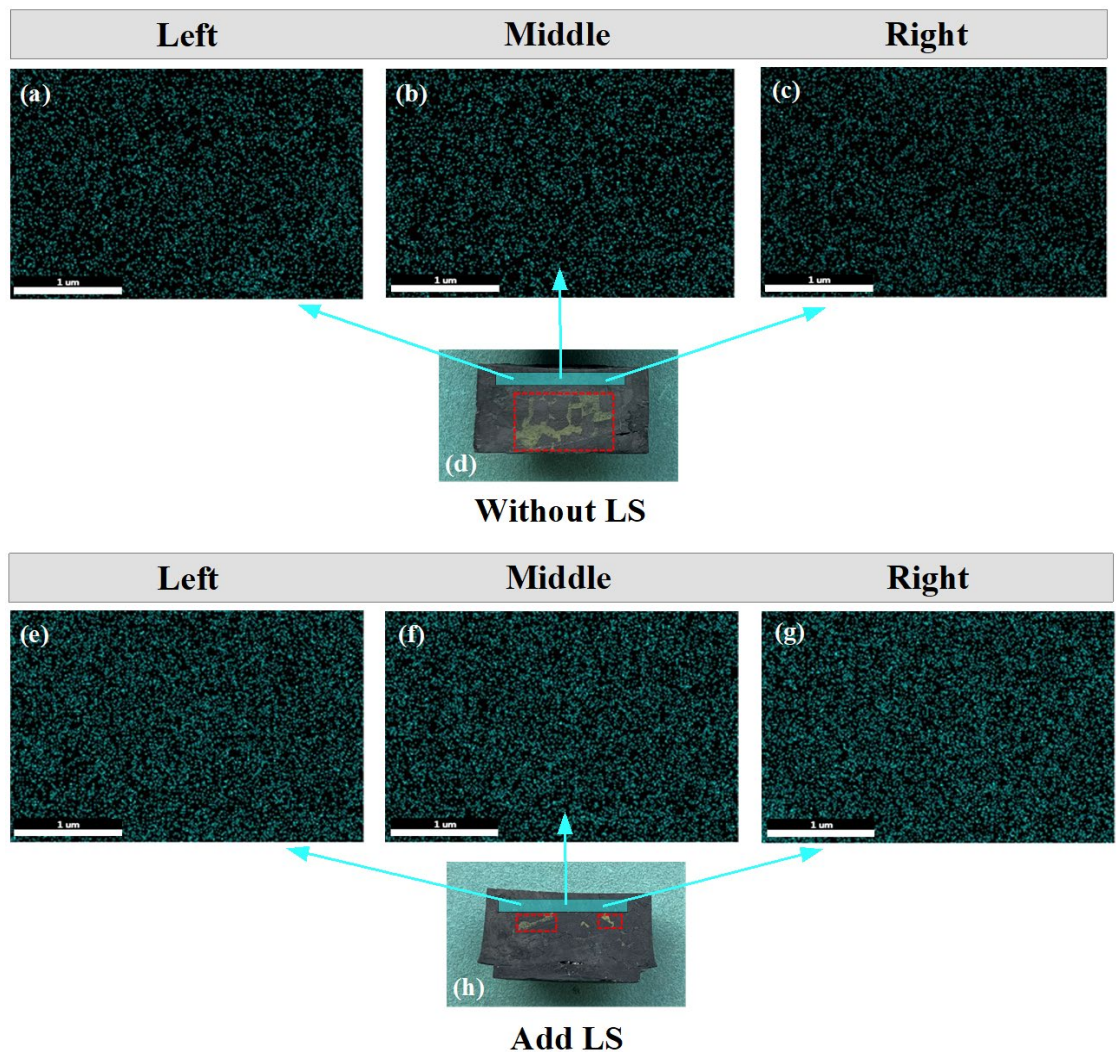

Fig. 11. Cross-section photographs of non-textured GdBCO samples and EDS mappings for Gadolinium ions. (a)-(d) belong to non-LS sample, (e)-(h) belong to LS sample. The red dashed boxes show green Gd211 precursor powders in the matrix.

## 4. Conclusion

In summary, a LS+LT$_{max}$ growth method has been employed for the fabrication of GdBCO single-grain superconductors. This method involves the incorporation of Y123 liquid source (LS) and the implementation of low maximum processing temperature (LT$_{max}$). The maximum processing temperature ($T_{max}$) is dramatically reduced by LS, even reaching peritectic decomposition temperature. We conducted a comparative analysis of the LS+LT$_{max}$ single grains and the Conventional single grains, employing morphology, superconducting properties and microstructure analysis as evaluative metrics. Our study indicates that the LS+LT$_{max}$ single grains exhibit a greater tendency towards perfect growth and excellent superconducting properties. This can be attributed that the LS+LT$_{max}$ growth method can uniform distribution of RE$^{3+}$ ions as well as RE211 particles, and the inhibition of RE211 coarsening resulting in more effective pining properties. Furthermore, our study also demonstrates that the pinning effect at the 123/211 interfaces is not the only contributing factor to the pinning at grain boundaries, and other pinning centers is also present. Last but not least, a scientific comprehension of the LS+LT$_{max}$ growth method and its related mechanisms can stimulate research extended to other oxide systems. Meantime, the remarkable performance of the LS+LT$_{max}$ growth method for material development is likely to make a breakthrough in engineering applications.

## Acknowledgments

This work was supported by the National Natural Science Foundation ((No. 52172271), the National Key R&D Program of China (No. 2022YFE03150200), and the Strategic Priority Research Program of the Chinese Academy of Sciences (No. XDB25000000).